\documentclass[]{aa}  

\usepackage{natbib}
\usepackage{amsmath}
\usepackage{txfonts}
\usepackage{bm}
\bibpunct{(}{)}{;}{a}{}{,} 

\usepackage{color}
\usepackage{graphicx}

\usepackage{txfonts}
\usepackage[draft]{hyperref}
\newcommand{\beq}{\begin{equation}}
\newcommand{\eeq}{\end{equation}}
\newcommand{\bea}{\begin{eqnarray}}
\newcommand{\eea}{\end{eqnarray}}

\newcommand{\cm}{\mathcal}

\definecolor{azul}{rgb}{0,0,.8}
\definecolor{rojo}{rgb}{1,0,0}
\definecolor{verde}{rgb}{0,.5,0}
\definecolor{violeta}{rgb}{.5,.0,1}
\definecolor{gris}{rgb}{.5,.5,.5}
\definecolor{marron}{rgb}{.4,.1,0}
\definecolor{naranja}{rgb}{1,.5,0}
\definecolor{bordo}{rgb}{.5,0,.2}
\definecolor{rojo}{rgb}{1,0,0}

\begin{document}

\title{Rayleigh scattering from hydrogen atoms including resonances and high photon energies}

\titlerunning{Rayleigh scattering from hydrogen atoms}

\author{Ren\'e D. Rohrmann \and Mat\'ias Vera Rueda} 
 
\institute{Instituto de Ciencias Astron\'omicas, de la Tierra y del
Espacio (CONICET-UNSJ), Av. Espa\~na 1512 (sur), 5400 San Juan, Argentina}

\abstract{
The nonrelativistic cross section from Rayleigh scattering by hydrogen atoms in the ground state was calculated over a wide range of photon energies ($< 0.8$~keV). Evaluations were performed in terms of the real and imaginary components of the atomic polarizability. The sum over intermediate states that characterizes this second-order radiative process was performed using exact analytic expressions for oscillator strengths of bound and continuum states. 
Damping terms associated with the finite lifetimes of excited states and their splitting into two fine-structure levels ($p_{1/2}$ and $p_{3/2}$) are taken into account in resonance cross sections.
Fitting formulas required for cross-section evaluation are presented for incident photon energy i) redward of the first resonance (Lyman-$\alpha_{1/2}$), ii) in the spectral region corresponding to resonances (for an arbitrary number of them), and iii) above the ionization threshold.
} 


\maketitle
%

\section{Introduction}\label{s:intro}

Rayleigh scattering is potentially relevant to several areas of astronomical spectroscopy including cool stars, star-forming regions, mass-losing stars, exoplanets, the circumgalactic medium, and cosmology. This process significantly affects opacity and emission of the monochromatic radiation field in stellar atmospheres at temperatures of a few thousand Kelvin \citep{marigo2009}. It also provides a diagnostic tool for determining geometrical parameters of double stellar systems containing a giant star \citep{isliker1989, gonzalez2003, skopal2012}, and the scale height and composition of exoplanetary atmospheres \citep{lecavelier2008, sing2015, dragomir2015}. 
Rayleigh scattering is especially important in spectroscopy for gaseous nebulae and emission regions of active galactic nuclei \citep{nussbaumer1989, ferland2017}.
In addition, scattering of light from isolated hydrogen atoms has an impact on the cosmic microwave background anisotropies.
The coupling of photons to neutral hydrogen through Rayleigh scattering during the recombination epoch may have had significant effects on the microwave background fluctuations' spectrum, providing in this way information about the formation of atoms in the early Universe \citep{scheuer1965, gunn1965, peebles1970, yu2001, bach2015, alipour2015, beringue2021}.

A coherent scattering cross section including resonances is required for radiation hydrodynamics simulations of late-type stars and accretion disks irradiated by the central star \citep{hayek2010, hirose2022}. 
It is an important ingredient in self-consistent magneto-hydrodynamical models of solar-type atmospheres which include a chromosphere \citep{hansteen2007, hayek2010}, and in detailed hydrodynamical models of giant stars \citep{collet2008, hayek2011}. 
Scattering resonances of the Lyman series may affect the temperature structure in the upper atmospheres of cool stars. Specifically, coherent line scattering reduces the temperature by several hundred degrees in the high atmosphere of solar-type stars \citep{hayek2010, hayek2011}. This effect could be even the most severe in the low-density giant star atmospheres, resulting in a significantly steeper temperature mean gradient.
On the other hand, resonant scattering processes affect the transfer of stellar radiation in protoplanetary disks and have implications on their photochemistry \citep{neufeld1991, bethell2011, heays2017}.

The present work is motivated by the incompleteness in the compiled set of theoretical Rayleigh cross sections from atomic hydrogen, which is typically limited to wavelengths redward of the Lyman-$\alpha$ resonance \citep{mittleman1962, isliker1989, lee2004, lee2005, fisak2016, colgan2016, hirose2022}.
An exact theoretical description of the Rayleigh scattering by one-electron systems was developed by \citet{gavrila1967}. However, his calculations partially covered only a few resonances and  they did not consider damping effects, yielding unphysical singularities. Scattering cross sections from atomic hydrogen have also been calculated by other authors \citep{nussbaumer1989, sadeghpour1992, mcnamara2018}, but their results have only been presented in a graphical form or over a limited range of photon frequencies and, therefore, they are not useful for computing purposes. In addition, such evaluations neglected fine-structure energy shifts due to relativistic and spin-orbit interaction effects.

The aim of the present work is to expand on the availability of  the Rayleigh cross section of neutral hydrogen atoms in the high-photon-energy regime, including an arbitrary number of resonances with excited bound states and taking into account the fine structure and finite lifetimes of such states. Current evaluations correspond to an isolated atom at rest and, therefore, they do not include perturbations arising from collisions with other atoms or Doppler shifts either due to translational motion of the radiating atom.

Rayleigh scattering represents a second-order photon-electron process in the Kramers-Heisenberg dispersion theory \citep{kramers1925, waller1929, chen2007}. The cross section for photons scattered by $1s$ bound electrons can be computed using the complex atomic polarizability (Bonin \& Kresin 1997). Fundamental properties of polarizability and scattering process were first treated by Placzek (1934).
Real and imaginary contributions to polarizability are expressed as sums of allowed dipole transitions to intermediate $np$ states, whose well-known oscillator strength values allow for an exact analytic evaluation \citep{penney1969}.
The imaginary part of polarizability has two terms related through the optical theorem to the spectral line absorptions and the photoionization process. 
Appropriate expressions for the scattering resonance result when level broadening effects are considered (Heddle 1964). 

The nonrelativistic dipole approximation adopted here is valid for photon energy much less than $2/\alpha \approx 275$~Rydberg (Ry) \citep{bethe1957}, with $\alpha=1/137.036$ being the fine-structure constant.
In this energy regime, relativistic effects on the dynamical polarizability can be neglected \citep{johnson1968, thu1996, zapryagaev2011}. For instance, the leading relativistic correction to the static dipole polarizability (nonrelativistic value of $9/2$~bohr$^3$), that is to say in the zero-photon-frequency  limit, is $\frac{14}3\alpha^2\approx 1.81\times10^{-6}$ \citep{kaneko1977}. However, corrections to the dynamic dipole polarizability become substantial near resonance peaks due to the corresponding energy shifts and splitting of bound states. 
Therefore, a semirelativistic model that accounts for the fine-structure effects is sufficient for calculating the cross section of Rayleigh scattering at energies both below and above the ionization threshold.

In Section \ref{s:frame}, we briefly describe the elastic photon scattering from isolated hydrogen atoms in the ground state through the complex polarizability and the use of transition oscillator strengths. We consider cross section corresponding to unpolarized incident radiation and outgoing radiation averaged over all directions. Sect. \ref{s:results} shows the results from numerical calculations done in the infinite level lifetime approximation and neglecting the fine structure. Sect. \ref{s:Gamma} is devoted to the analysis of the cross section in the neighborhood of resonances when the natural broadening of excited bound states and the fine structure are taken into account. 
In Sect. \ref{s:fits} we give functional forms which fit the atomic polarizability.  Sect. \ref{s:cross} presents some evaluations of the Rayleigh cross section. Conclusions are given in Sect. \ref{s:concl}.

\section{Evaluation of the atomic polarizability}\label{s:frame}

The nonrelativistic cross section of Rayleigh scattering of unpolarized light by a nonoriented hydrogen atom in the ground state, expressed in atomic units (bohr$^2$), is given by
\beq \label{Ra1}
\sigma_\text{Ra}(\epsilon) = \frac{\pi}{6}\alpha^4\epsilon^4 
               \left|\alpha_\text{pol}(\epsilon)\right|^2,
\eeq
with $\epsilon$ being the photon energy in Rydberg units ($R=13.60569$~eV), and $\alpha_\text{pol}(\epsilon)$ being the dynamic polarizability measured in bohr$^3$. 
The polarizability is constituted by real and imaginary parts
\beq\label{alphaP}
\alpha_\text{pol}=\alpha_\text{R}+i \alpha_\text{I},\hskip.4in
|\alpha_\text{pol}|^2=\alpha_\text{R}^2+\alpha_\text{I}^2.
\eeq
They can be expressed in terms of the oscillator strengths of bound states ($f_{1n}$) and the continuum ($df_{1k}/d\epsilon$), 
\beq\label{alphaR}
\alpha_\text{R}^0(\epsilon) = 4\left[
 \sum_{n=2}^\infty\frac{f_{1n}}{\epsilon_{1,n}^2-\epsilon^2}
 +\cm{P}\int_{1}^\infty\frac{df_{1k}/d\epsilon_{1,k}}{\epsilon_{1,k}^2-\epsilon^2}
   d\epsilon_{1,k} \right],
\eeq
\beq\label{alphaI}
\alpha_\text{I}^0(\epsilon) =  \sum_{n=2}^\infty\frac{2\pi }
    {\epsilon_{1,n}} f_{1n} \delta\left(\epsilon_{1,n}-\epsilon\right)
   + \frac{2\pi}{\epsilon}\left(\frac{df_{1k}} {d\epsilon}\right),
\eeq
where $\cm{P}$ denotes the principal value of the integral, $\delta(x)$ is the Dirac function, and the superscript zero means that bound states are assumed to have infinite lifetimes. In addition, $\epsilon_{1,n}$ and $\epsilon_{1,k}$ are the energies (measured in Rydberg) over the ground state $n'=1$
\beq\label{Enn}
\epsilon_{n',n}=\frac{1}{n'^2}-\frac{1}{n^2},\hskip.3in 
\epsilon_{n',k}=\frac{1}{n'^2}+\frac{1}{k^2},
\eeq
where $n$ ($=2,3,\dots,\infty$) represents the main quantum number of bound states, and $k$ is a positive real number ($0<k<\infty$) labeling states in the continuum.

The real part of the polarizability ($\alpha_\text{R}$) is associated with the refractive index of the medium and can have positive or negative values.
The imaginary part of the polarizability  ($\alpha_\text{I}$) is a positive quantity related to the absorption cross section through the optical theorem. In the continuum ($\epsilon>1$), it corresponds to the photoionization cross section given in bohr$^2$ by
\beq\label{photo}
\sigma_\text{Ph}(\epsilon) = 2\pi\alpha\, \epsilon \, \alpha_\text{I}.
\eeq
As a reference, the Thomson cross section in the same units is
\beq\label{thomson}
\sigma_\text{Th} = \frac{8\pi}{3}\alpha^4.
\eeq

Exact expressions for the mean oscillator strength of transitions $1s\rightarrow np$ and  $1s\rightarrow kp$ are well known \citep{sugiura1927,menzel1935},\footnote{It is worth noting that, for the ground state $n'=1$, the averaged oscillator strength $f_{n'n}$ is equal to that from the sublevel transition, $f_{n's,np}$.}
\beq\label{f1n}
 f_{1n}=\frac{256}{3}\frac{n^5(n-1)^{2(n-2)}}{(n+1)^{2(n+2)}},
\eeq
\beq\label{dfde}
\frac{df_{1k}}{d\epsilon_{1,k}}=\frac{128}{3}\frac{k^8}
 {(1+k^2)^{4}} \frac{\exp\left[-4k\arctan(1/k)\right]}
  {1-\exp\left[-2\pi k\right]} .
\eeq
The oscillator strength has analytic continuation through the ionization threshold \citep{fano1968},
\beq \label{border1}
\lim_{n\rightarrow \infty} \left(\frac{n^3}{2} f_{1n} \right) 
  = \lim_{k\rightarrow\infty} \left(\frac{df_{1k}}{d\epsilon_{1,k}}\right)
=\frac{128}{3\exp(4)}\approx 0.78146725925.
\eeq
Moreover, the asymptotic behavior of the oscillator strength is expressed by the series
\bea \label{border2}
\frac{n^3}2f_{1n} &=&
 \frac{128}{3 \exp(4)} \left[ 1 +\frac{8}{3n^2} +\frac{214}{45n^4}
  +\frac{20192}{2835n^6} +\frac{411683}{42525n^8}  
 \right.\cr &+&\left.
  \frac{17369584}{1403325n^{10}} +\cm{O}\left(n^{-12}\right) \right].
\eea
The same expression is valid for $df_{1k}/d\epsilon_{1,k}$ with the substitution $n^2=-k^2$ ($n=ik$) on the right-hand side of Eq. (\ref{border2}). Direct use of Eqs. (\ref{f1n}) and (\ref{dfde}) yields numerical errors for high quantum numbers, so we employed Eq. (\ref{border2}) to evaluate $f_{1n}$ and $df_{1k}/d\epsilon_{1,k}$ for $n,k\gg 1$. 

The principal value integral in (\ref{alphaR}) reduces to a regular integral for $\epsilon<1$ and the imaginary pole term can be ignored. For $\epsilon>1$, Cauchy principal value reads as follows:
\beq\label{ppal}
 \cm{P}\int_{1}^\infty \dots = \lim_{\delta\rightarrow 0} \left(
\int_{1}^{\epsilon-\delta}\dots+\int_{\epsilon+\delta}^\infty\dots \right).
\eeq
In practice, the evaluation of this term is split into three subdomains:
\beq\label{ppal2}
 \cm{P}\int_{1}^\infty \dots=  \int_{1}^{\epsilon-\delta}\dots
 + \int_{\epsilon-\delta}^{\epsilon+\delta}\dots
 + \int_{\epsilon+\delta}^\infty\dots, \hskip.3in (\delta\ll \epsilon).
\eeq
The second-term integrand on the right-side of Eq. (\ref{ppal2}) is approximated by its Laurent expansion, where the odd terms about $\epsilon$ are removed.
Each even term in $\epsilon$ can be analytically integrated yielding the following series:
\bea
\int_{\epsilon-\delta}^{\epsilon+\delta}\frac{df_{1k}/d\epsilon_{1,k}}
 {\epsilon_{1,k}^2-\epsilon^2}d\epsilon_{1,k} 
 = \left(-f_\epsilon+2\epsilon f^{(1)}_\epsilon\right)
    \frac{\delta}{2\epsilon^2}
  \hskip.1in \cr
  + \left(-3f_\epsilon+6\epsilon f^{(1)}_\epsilon -6\epsilon^2 f^{(2)}_\epsilon
  +4\epsilon^3 f^{(3)}_\epsilon\right)\frac{\delta^3}{72\epsilon^4}
\hskip.1in \cr
  + \left(-15f_\epsilon +30\epsilon f^{(1)}_\epsilon 
   -30\epsilon^2 f^{(2)}_\epsilon  +20\epsilon^3 f^{(3)}_\epsilon 
    \right. 
\hskip.1in \cr
  \left. -10\epsilon^4 f^{(4)}_\epsilon +4\epsilon^5 f^{(5)}_\epsilon\right)
\frac{\delta^5}{2400\epsilon^6}  
 +\cm{O}\left(\delta^7\right),\hskip.1in \cr
\eea
where $f^{(l)}_\epsilon$ is evaluated with (\ref{dfde}) as follows:
\beq
f^{(l)}_\epsilon\equiv 
    \frac{\partial^l (df_{1k}/d\epsilon)}{\partial \epsilon^l}.
\eeq
The first and third integrals on the right-hand side of (\ref{ppal2}) were calculated in a standard way using Gaussian quadratures.

\section{Results for infinite lifetimes} \label{s:results}

\begin{figure}
\includegraphics[width=.51\textwidth]{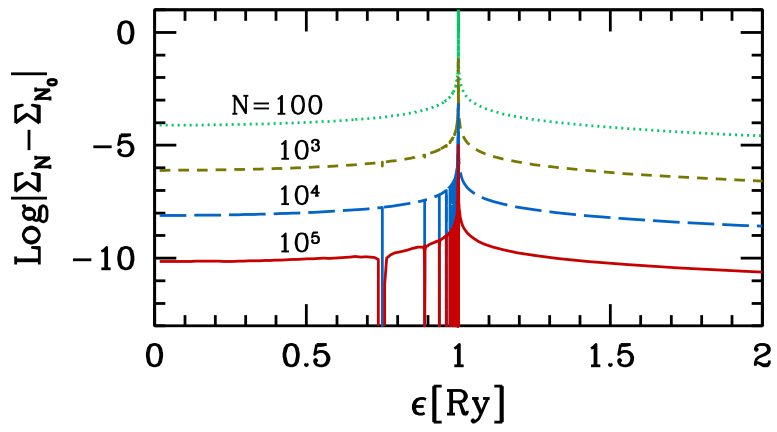}
\caption{Accuracy in the evaluation of bound states' contribution to real polarizability, as a function of the photon energy and for different numbers of sum terms in (\ref{sigmaN}). The reference values $\Sigma_{N_0}$ correspond to $N_0=10^6$. Vertical lines are located on resonances and show a fast convergence effect.} \label{f:fdis}
\end{figure}

Expressions (\ref{alphaR}) and (\ref{alphaI}) provide high-precision values of $\alpha_\text{pol}$ far away from resonance cores. This section shows the results of their evaluation. In practice, the sum over bound states in (\ref{alphaR}) is truncated to some upper number $N$:
\beq\label{sigmaN}
\Sigma_N\equiv\sum_{n=2}^N\frac{f_{1n}}{\epsilon_{1,n}^2-\epsilon^2}.
\eeq
Convergence in the evaluation of $\alpha_\text{R}^0(\epsilon)$ was reached by increasing $N$ and the number of points in the quadratures. 
Fig. \ref{f:fdis} shows the sensitivity of (\ref{sigmaN}) to $N$. Precision in the sum increases roughly two orders of magnitude for each one-order increase in the number of bound states.

Determining the accuracy by using the use of spectral distribution of oscillator strengths can be done through two simple tests: i) the Thomas-Reiche-Kuhn f-sum rule 
\beq
\sum_{n=2}^\infty f_{1n} 
 +\int_{1}^\infty\frac{df_{1k}}{d\epsilon_{1,k}}d\epsilon_{1,k} =1,
\eeq
and ii) the static polarizability value which is exactly know \citep{wentzel1926, waller1926, epstein1926}\footnote{ The exact value for the static polarizability of hydrogen atoms can also be obtained by the so-called Dalgarno-Lewis method \citep{dalgarno1955,dalgarno1960}.}: 
\beq
 \alpha_\text{pol}(0)=4.5~\text{bohr}^3.
\eeq
As shown in Table \ref{T.1}, both tests can be verified within machine precision. Here, an upper quantum number $N=10^6$ is adopted. 
\begin{table}
\caption{Contributions to the $f$-sum rule and  static polarizability 
$\alpha_\text{pol}(0)$ coming from bound and continuum states.
\label{T.1}}
\setlength{\tabcolsep}{3pt}
\begin{small}
\begin{tabular}{cccc}\hline
states  & $f$-sum  & $\alpha_\text{pol}(0)$[bohr$^3$] & $\alpha_\text{pol}(0)$[\%]\\
\hline 
 discrete \hskip.2in &   $0.5650041506$ &  $3.66325789028$ & $81.4057308951$ \\
 continuum\hskip.1in &   $0.4349958493$ &  $0.83674210969$ & $18.5942691042$  \\
  all\hskip.2in      &   $0.9999999999$ &  $4.49999999997$ & $99.9999999993$ \\
\hline
\end{tabular}
\end{small}
\end{table}

\begin{figure}
\includegraphics[width=.5\textwidth]{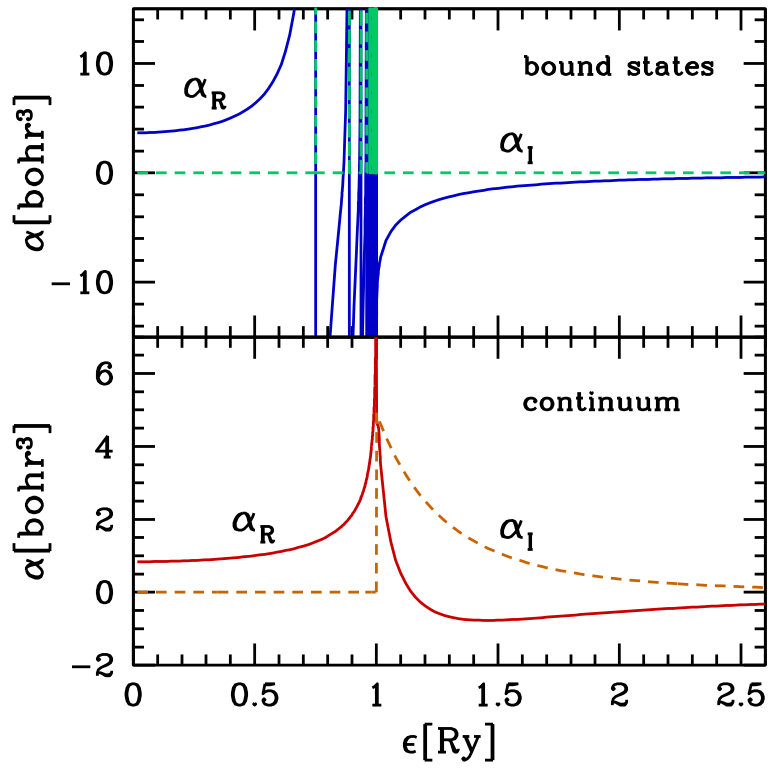}
\caption{Variation of the real and imaginary parts of the polarizability due to contributions from bound and continuum states.}
\label{f:fpolRI}
\end{figure}
Fig. \ref{f:fpolRI} shows different contributions to the dynamical polarizability as functions of the photon energy. The value $\epsilon=1$  corresponds to the photoelectric threshold for transitions from the $1s$ state. 
When the natural broadening of the levels is neglected, real and imaginary parts of the polarizability due to bound states become singulars over an infinite sequence of resonances located at energies $\{\epsilon_{1,n}\}$ ($n=1,2,\dots,\infty$), which are distributed at $0.75\le \epsilon\le 1$ and accumulate on the photoionization edge (upper panel of Fig. \ref{f:fpolRI}). In this approach, the contribution of bound states to $\alpha_\text{I}^0(\epsilon)$ vanishes for all energies outside of resonances (i.e., $\forall~\epsilon\ne\epsilon_{1,n}$) according to the set of Dirac functions in Eq. (\ref{alphaI}).

Contributions of the continuum to polarizability exhibit simple forms (lower panel of Fig. \ref{f:fpolRI}). The real part increases monotonically with energy up to $\epsilon=1$, where it diverges. For energies higher than $1.144210$~Ry, $\alpha_\text{R}$ from the continuum becomes negative and reaches a minimum value of $-0.7712916$~bohr$^3$ for $\epsilon\approx 1.4540$. On the other hand, the imaginary part mimics  -- with a multiplicative factor proportional to $\epsilon$ -- the behavior of the photoabsorption cross section, according to Eq. (\ref{photo}).

\begin{figure}
\includegraphics[width=.5\textwidth]{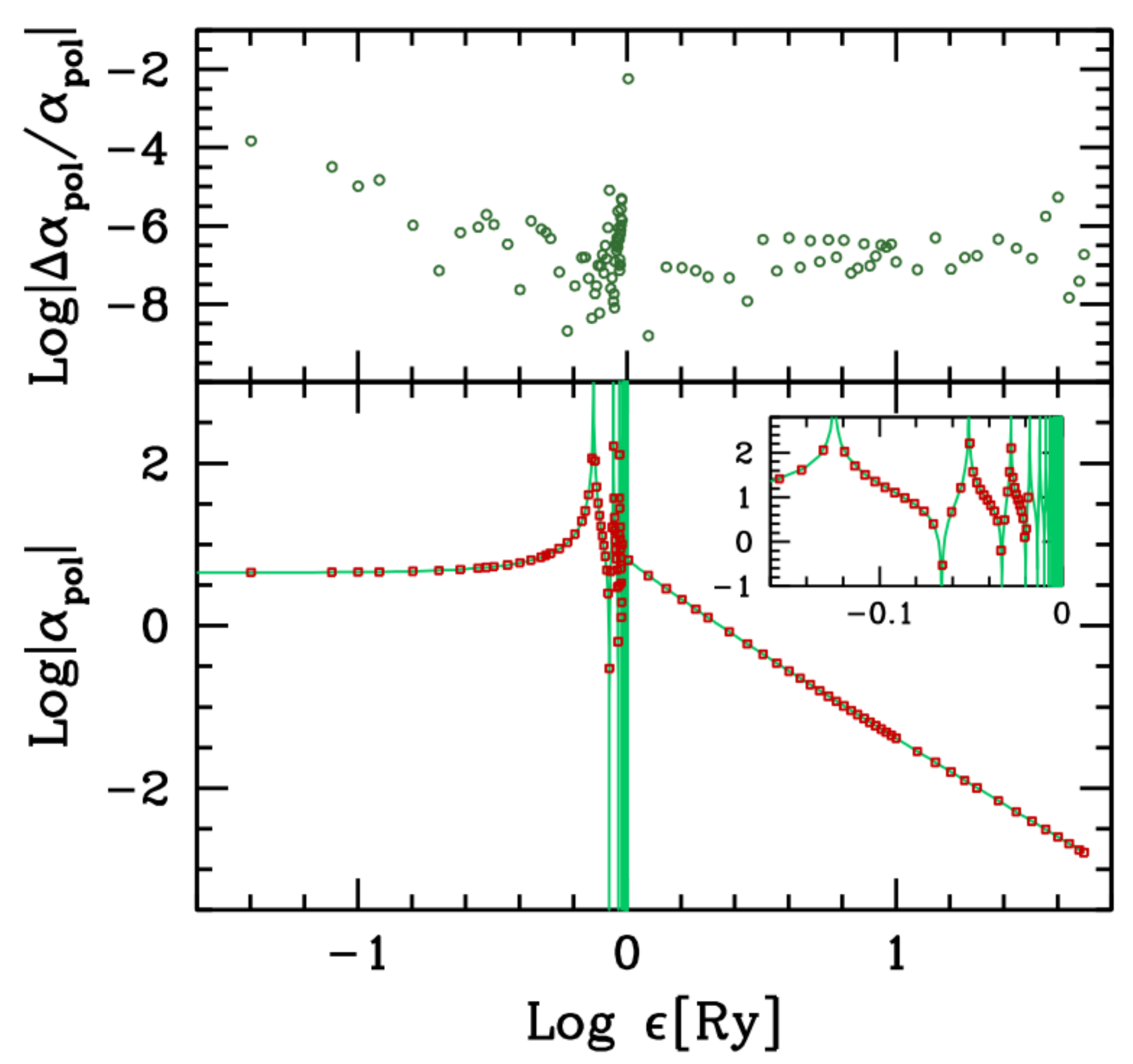}
\caption{Dynamic polarizability. Lower panel: Comparison between the values calculated in the present work (line) and the results of \citet{gavrila1967} (symbols), as a function of the light energy. Upper panel: Absolute relative errors in a logarithmic scale.}
\label{f:fpol}
\end{figure}
The absolute magnitude of $\alpha_\text{pol}$ as a function of $\epsilon$ is illustrated in logarithmic scales in the lower panel of Fig. \ref{f:fpol}, where our results (solid line) are compared with those from \citet{gavrila1967} which are represented by symbols. 
Relative differences are plotted in the upper panel of Fig.  \ref{f:fpol}.
In the limit $\epsilon\rightarrow 1^+$, contributions to $\alpha_\text{R}$ from discrete states and the continuum diverge with opposite signs (see Fig. \ref{f:fpolRI}),
but they compensate for each other in such a way that total $\alpha_\text{R}$ remains finite. Therefore, observed divergences  of $\alpha_\text{pol}$ in our evaluations are only produced by resonances, and as a consequence of neglecting the level broadening.

\section{Fine-structure and damping effects} 
\label{s:Gamma}

When the effects of a fine structure and finite lifetimes of excited bound states are considered, real and imaginary parts of polarization take the forms
\beq\label{alphaRD}
\alpha_\text{R}(\epsilon)=4\left[\sum_{nj}
\frac{f_{1,nj}(\epsilon_{1,nj}^2-\epsilon^2)}
       {(\epsilon_{1,nj}^2-\epsilon^2)^2+\epsilon^2\Delta_n^2}
 +\, \cm{P}\int_{1}^\infty\frac{df_{1k}/d\epsilon_{1k}}
      {\epsilon_{1,k}^2-\epsilon^2}d\epsilon_{1k}\right],
\eeq
\beq\label{alphaID}
\alpha_\text{I}(\epsilon)=  \sum_{nj}\frac{4 f_{1,nj}\epsilon\Delta_n}
                {(\epsilon_{1,nj}^2-\epsilon^2)^2+\epsilon^2\Delta_n^2}
  + \frac{2\pi}{\epsilon}\left(\frac{df_{1k}}{d\epsilon}\right),
\eeq
where $\Delta_n$ is the natural breadth of the level $n$ and $j$ refers to the two components of each state $np$ ($np_{1/2}$ and $np_{3/2}$). Absorption oscillator strengths of fine-structure transitions are given by the following \citep{wiese2009}:
\beq\label{f1nj}
f_{1,nj=1/2}= \frac13 f_{1n},\hskip.5in
f_{1,nj=3/2}= \frac23 f_{1n}.
\eeq
The natural breadth is basically the same for $np_{1/2}$ and $np_{3/2}$ levels, and it can be written in Rydberg units as
\beq\label{D_n}
\Delta_n= \frac{\hbar}{R}\Gamma_n,\hskip.5in
\Gamma_n=\sum_{n'=1}^{n-1}A_{nn'},
\eeq
with $\Gamma_n$[s$^{-1}$] being the total probability rate of spontaneous decay from $n$ to any lower level, and $A_{nn'}$[s$^{-1}$] is the Einstein coefficient,
\beq\label{Ann}
A_{nn'}=\frac{\alpha^3 R}{\hbar}\frac{g_{n'}}{g_n}\epsilon_{n',n}^2 f_{n'n}
= 8.03250\times10^9[\text{s}^{-1}]\frac{g_{n'}}{g_n}\epsilon_{n',n}^2 f_{n'n},
\eeq
with $g_n$ being the statistical weight of the level $n$. 
In practice, we calculated $\Delta_n$ using accurate $A_{nn'}$ values compiled by \citet{wiese2009}, which expand $n'\le 19$ and $n\le 20$ with an uncertainty of less than 0.3\%. 
A precise (within data errors) $\Delta_n$ fitting expression for $n\ge 2$ is given by
\beq\label{D_ajus}
\Delta_n^\text{fit} =\frac{\alpha^3}{n^5}\left(-0.187+2.915\ln n\right).
\eeq
Table \ref{T.3} shows calculated values for natural breadths and probabilities of  a spontaneous transition for a selection of levels.
Errors of $\Delta_n^\text{fit}$ are lower than 0.2\% for $n\ge 2$. As a reference, evaluations of spontaneous decay probabilities from the well-known Kramers approximation, 
\beq
A^\text{K}_{nn'}=\frac{32}{3\pi\sqrt{3}}\frac{\alpha^3 R}{\hbar}\frac{1}{n'^3 n^5 \epsilon_{n'n}},
\eeq
yield $\Delta_n$ values with errors between 14\% and 40\% for $n\le 20$.
\begin{table}
\caption{Selection of transition probabilities and natural level breadths.
We note that $\chi(\text{fit})$ and $\chi(\text{K})$ express the relative errors of the $\Delta_n$ evaluation using Eq. (\ref{D_ajus}) and the Kramers approximation, respectively. Numbers in brackets indicate powers of ten.\label{T.3}}
\setlength{\tabcolsep}{3pt}
\begin{small}
\begin{tabular}{rlllcc}  
\hline 
  $n$ & $\hskip.15in\epsilon_{1,n}$[Ry] &  $\hskip.1in\Gamma$[1/s]
      & $\hskip.1in\Delta_n$[Ry] & $\chi(\text{fit})$(\%) &
        $\chi(\text{K})$(\%)  \\\hline
  $2$ & $0.75$         & $4.69860(8)$  & $2.27307(-8)$  & $+2.04$ & $-39.6$ \\
  $3$ & $0.8888888889$ & $9.98520(7)$  & $4.83061(-9)$  & $+0.17$ & $-31.4$ \\
  $4$ & $0.9375$       & $3.01903(7)$  & $1.46054(-9)$  & $-0.13$ & $-27.1$ \\
  $5$ & $0.96$         & $1.15555(7)$  & $5.59029(-10)$ & $-0.20$ & $-24.4$ \\
  $6$ & $0.9722222222$ & $5.19199(6)$  & $2.51177(-10)$ & $-0.19$ & $-22.5$ \\
  $7$ & $0.9795918367$ & $2.61709(6)$  & $1.26609(-10)$ & $-0.17$ & $-21.0$ \\
  $8$ & $0.984375$     & $1.43796(6)$  & $6.95652(-11)$ & $-0.15$ & $-19.9$ \\
  $9$ & $0.9876543210$ & $8.44833(5)$  & $4.08711(-11)$ & $-0.12$ & $-19.0$ \\
 $10$ & $0.99$         & $5.23645(5)$  & $2.53327(-11)$ & $-0.09$ & $-18.3$ \\
 $11$ & $0.9917355372$ & $3.39069(5)$  & $1.64034(-11)$ & $-0.07$ & $-17.6$ \\
 $12$ & $0.9930555556$ & $2.27687(5)$  & $1.10150(-11)$ & $-0.05$ & $-17.1$ \\
 $13$ & $0.9940828402$ & $1.57668(5)$  & $7.62761(-12)$ & $-0.02$ & $-16.7$ \\
 $14$ & $0.9948979592$ & $1.12094(5)$  & $5.42283(-12)$ & $-0.01$ & $-16.2$ \\
 $15$ & $0.9955555556$ & $8.15308(4)$  & $3.94427(-12)$ & $+0.01$ & $-15.9$ \\
 $16$ & $0.99609375$   & $6.04947(4)$  & $2.92659(-12)$ & $+0.03$ & $-15.5$ \\
 $17$ & $0.9965397924$ & $4.56826(4)$  & $2.21002(-12)$ & $+0.04$ & $-15.2$ \\
 $18$ & $0.9969135802$ & $3.50397(4)$  & $1.69514(-12)$ & $+0.05$ & $-14.9$ \\
 $19$ & $0.9972299169$ & $2.72545(4)$  & $1.31851(-12)$ & $+0.06$ & $-14.7$ \\
 $20$ & $0.9975$       & $2.14670(4)$  & $1.03853(-12)$ & $+0.08$ & $-14.5$ \\
\hline
\end{tabular}
\end{small}
\end{table}

Due to relativistic corrections and spin-orbit interaction, each $np$ state splits into two levels with energies \citep{sobelman1979}
\beq\label{efine}
\epsilon_{np_{1/2}} =-\frac1{n^2}+ \frac{\alpha^2}{n^3}\left(\frac{3}{4n}-1\right),\hskip.05in
\epsilon_{np_{3/2}} =-\frac1{n^2}+ \frac{\alpha^2}{n^3}\left(\frac{3}{4n}-\frac12\right),
\eeq
with the zero-energy point in the continuum edge. The ground state ($1s$) remains single ($1s_{1/2}$), but its energy changes from $\epsilon_{1s}=-1$~Ry to 
\beq
\epsilon_{1s_{1/2}} =-1.0000133128~\text{Ry}.
\eeq
As a consequence, each resonance $1s$~--~$np$ splits into a doublet with energies
\beq\label{e1nj}
\epsilon_{1,nj}= \epsilon_{np_{j}} -\epsilon_{1s_{1/2}},
\hskip.2in \left(j=\frac12,\frac32\right)
,\eeq
which are slightly higher than that from Eq. (\ref{Enn}).
They are listed for $n\le 20$ in Table \ref{T.4} along with the corresponding transition wavelengths \citep[compare them with][]{kramida2010}.

\begin{table}
\caption{Resonances energies (Ry) and wavelengths (\AA) of transitions $1s_{1/2}$--$np_{1/2}$ and $1s_{1/2}$--$np_{3/2}$ for $2\le n\le 20$.
\label{T.4}}
\setlength{\tabcolsep}{3pt}
\begin{small}
\begin{tabular}{rccrr}  
\hline 
 $n$ & $\epsilon(1s_{1/2},np_{1/2})$ & $\epsilon(1s_{1/2},np_{3/2})$ 
     & $\lambda(1s_{1/2},np_{1/2})$ & $\lambda(1s_{1/2},np_{3/2})$ \\ \hline
  $2$ & $0.7500091526$ & $0.7500124808$ &
       $1215.67364461$ & $1215.66825001$ \\
  $3$ & $0.8889007225$ & $0.8889017087$ &
       $1025.72349971$ & $1025.72236178$ \\
  $4$ & $0.9375126368$ & $0.9375130528$ &
       $972.53767492$ & $972.53724335$ \\
  $5$ & $0.9600129507$ & $0.9600131637$ &
       $949.74381263$ & $949.74360190$ \\
  $6$ & $0.9722353193$ & $0.9722354426$ &
       $937.80419396$ & $937.80407506$ \\
  $7$ & $0.9796050110$ & $0.9796050886$ &
       $930.74897515$ & $930.74890140$ \\
  $8$ & $0.9843882186$ & $0.9843882706$ &
       $926.22640416$ & $926.22635523$ \\
  $9$ & $0.9876675669$ & $0.9876676034$ &
       $923.15105870$ & $923.15102457$ \\
 $10$ & $0.9900132636$ & $0.9900132902$ &
       $920.96378255$ & $920.96375778$ \\
 $11$ & $0.9917488127$ & $0.9917488328$ &
       $919.35210638$ & $919.35208784$ \\
 $12$ & $0.9930688395$ & $0.9930688549$ &
       $918.13006685$ & $918.13005260$ \\
 $13$ & $0.9940961302$ & $0.9940961424$ &
       $917.18127882$ & $917.18126764$ \\
 $14$ & $0.9949112537$ & $0.9949112634$ &
       $916.42983899$ & $916.42983006$ \\
 $15$ & $0.9955688534$ & $0.9955688613$ &
       $915.82451267$ & $915.82450541$ \\
 $16$ & $0.9961070504$ & $0.9961070569$ &
       $915.32969232$ & $915.32968634$ \\
 $17$ & $0.9965530949$ & $0.9965531003$ &
       $914.92000245$ & $914.91999748$ \\
 $18$ & $0.9969268843$ & $0.9969268889$ &
       $914.57696078$ & $914.57695660$ \\
 $19$ & $0.9972432223$ & $0.9972432262$ &
       $914.28684561$ & $914.28684205$ \\
 $20$ & $0.9975133064$ & $0.9975133098$ &
        $914.03929564$ &  $914.03929259$ \\
\hline
\end{tabular}
\end{small}
\end{table}

\begin{figure}
\includegraphics[width=.5\textwidth]{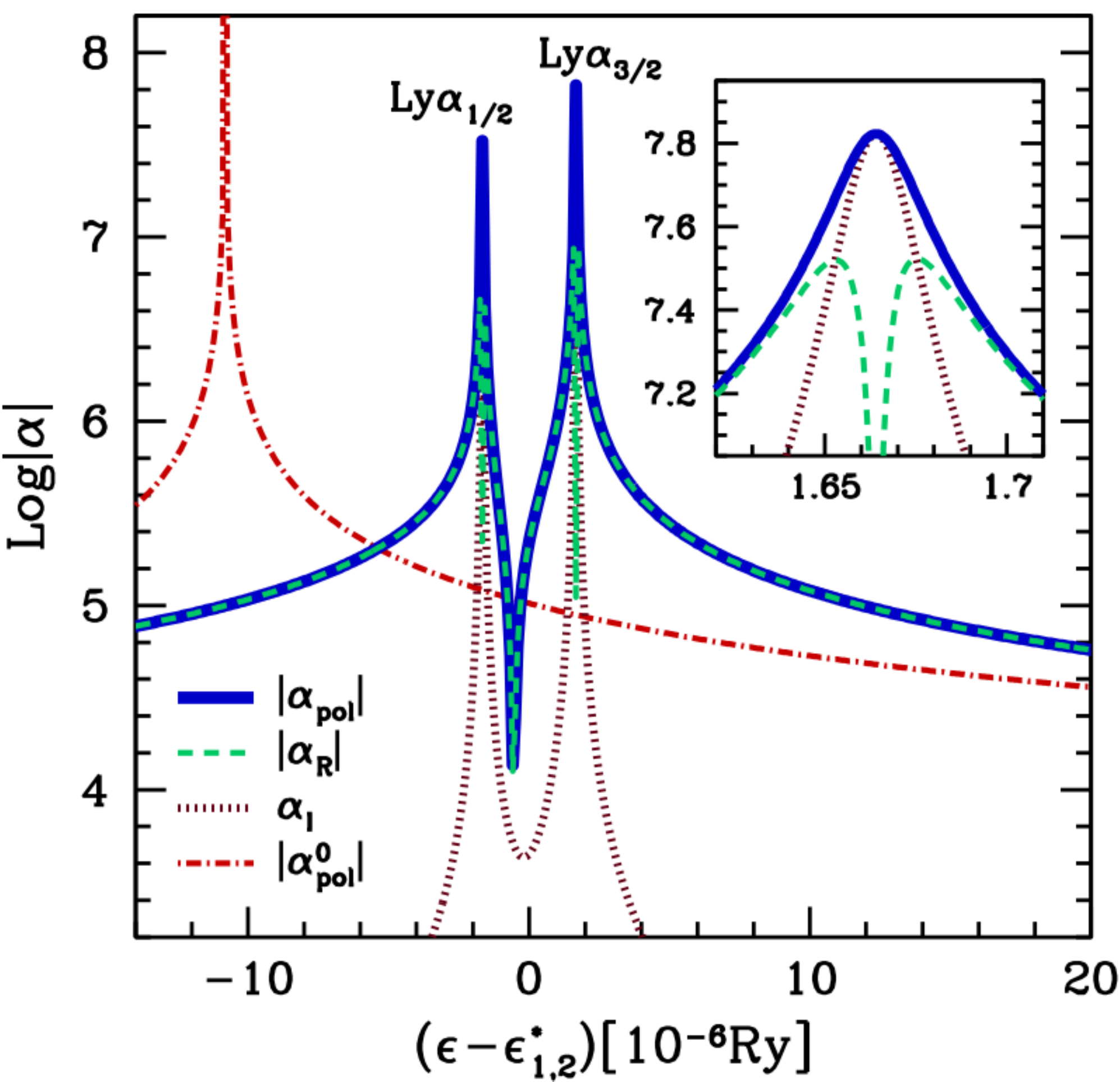} 
\caption{Polarizability in the Ly$\alpha_{1/2}$ ($1s_{1/2}$~--~$2p_{1/2}$) and Ly$\alpha_{3/2}$ ($1s_{1/2}$~--~$2p_{3/2}$) resonances (solid line). We note that $\epsilon^*_{1,2}$ denotes the mean energy of these transitions. Real and imaginary parts of the polarizability are detailed on the plot (dashed and dotted lines, respectively).  The inner graph shows details of the Ly$\alpha_{3/2}$ core. Evaluation of Lyman-$\alpha$ resonance without the effects of a fine structure and finite lifetime is represented by a dot-dashed line. 
}
\label{f:fLy1}
\end{figure}
Polarizability contributions (\ref{alphaRD}) and (\ref{alphaID}) were evaluated in the way described in Sect. \ref{s:frame}.  
They provide different results than those given by Eqs. (\ref{alphaP})-(\ref{alphaI}) in the neighborhood of each resonance.
The first pair of resonances occur around $\epsilon=0.75$ and correspond to Lyman-$\alpha_{1/2}$ and Lyman-$\alpha_{3/2}$ transitions. 
Fig. \ref{f:fLy1} shows the module of the polarizability with (solid line) and without (dot-dashed line) a fine structure and damping effects for these resonances. Resonance polarizability given by Eqs. (\ref{alphaR}) and (\ref{alphaI}) is redshifted in $1.0816\times 10^{-5}$~Ry (line center in $\epsilon_{1,n}=0.75$~Ry) and its real and imaginary contributions become singulars. On the contrary, imaginary polarizability $\alpha_\text{I}$ with damping and fine-structure effects (dotted line) presents sharp Lorentzian peaks centered at $\epsilon_{1,2p_{1/2}}$ and  $\epsilon_{1,2p_{3/2}}$, with a half width $\Delta_n$. The maximum values of $\left|\alpha_\text{pol}\right|$ coincide with the $\alpha_\text{I}$ peaks since the real part (dashed line) vanishes there. In fact, the real part of the polarizability tends to be antisymmetric about each resonance center (Fig. \ref{f:fLy1} shows its absolute value).

\section{Analytic fits} \label{s:fits}

In this section we provide expressions to evaluate the absolute value of the polarizability. As has been shown, the imaginary part of the polarizability is significant in a very small region around each resonance (Fig. \ref{f:fLy1}) and in the photoionization region due to the continuum contribution (Fig. \ref{f:fpolRI}). Therefore, for energies lower than the ionization threshold ($\epsilon<\epsilon_{1s_{1/2}}$) and outside resonance cores, the magnitude of the polarizability ($|\alpha_\text{pol}|$) is very well approximated by its real part
\beq\label{alphaPR}
\left|\alpha_\text{pol}(\epsilon)\right| = 
        \left|\alpha_\text{R}(\epsilon)\right|,
\hskip.2in
(\epsilon<\epsilon_{1s_{1/2}},\epsilon\ne\epsilon_{1,nj}).
\eeq

{\it Preresonance region} ($\epsilon\la \epsilon_{1,2}=0.75$~Ry). Redward of Lyman-$\alpha$, the polarizability is a wellbehaved monotonic function of $\epsilon$ and can be approximated with high precision (relative error less than $0.006\%$ at $\epsilon<0.7496$) by
\beq \label{aRpre}
  \alpha_\text{R}(\epsilon) = \frac{1}{1-s}
           \left( \frac{1.46486}{0.950713-\epsilon^{2.172}}
                 +\frac{1.66478}{{\epsilon^*_{1,2}}^2-\epsilon^2}\right),
\eeq
being $\epsilon^*_{1,2}=\frac12(\epsilon_{1,2~j=1/2}+\epsilon_{1,2~j=3/2})$ and
\beq \label{xi}
s = \left\{ \begin{array}{ll}\displaystyle
           0.0017\sin\left(8.2\epsilon^{1.33}\right)-0.000093
                        & (\epsilon\le \epsilon_a),\\
     [2ex]\displaystyle 
          -0.00163\sin\left(16.86\left|\epsilon-\epsilon_a\right|^{1.2}\right)
                        & (\epsilon_a <\epsilon < 0.73),\\
     [2ex]\displaystyle 
         0                & (0.73 <\epsilon < 0.745),\\
     [2ex]\displaystyle 
     -10^{-4.9+0.205(0.7501-\epsilon)^{-0.3}}   & (0.745 <\epsilon < 0.75),\\
\end{array} \right.
\eeq
$\epsilon_a=0.48083$ (for $s\equiv 0$, Eq. (\ref{aRpre}) has a precision of 0.2\%).

{\it Resonance region} ($\epsilon_{1,2}\la\epsilon<|\epsilon_{1s_{1/2}}|$). 
The real polarizability in the resonance region can be reasonably well approximate in the following way. In the neighborhood of a resonance $1s\leftrightarrow np$, the polarizability is well represented keeping only the contributions of  $1s_{1/2}$--$np_{1/2}$ and $1s_{1/2}$--$np_{3/2}$ transitions,
\beq\label{alphaRDn}
\alpha_\text{R}(\epsilon)= \frac{4f_{1,nj=1/2}(\epsilon_{1,nj=1/2}^2-\epsilon^2)}
       {(\epsilon_{1,nj=1/2}^2-\epsilon^2)^2+\epsilon^2\Delta_n^2}
+\frac{4f_{1,nj=3/2}(\epsilon_{1,nj=3/2}^2-\epsilon^2)}
       {(\epsilon_{1,nj=3/2}^2-\epsilon^2)^2+\epsilon^2\Delta_n^2},
\eeq
\beq\label{alphaIDn}
\alpha_\text{I}(\epsilon)= \frac{4 f_{1,nj=1/2}\epsilon\Delta_n}
         {(\epsilon_{1,nj=1/2}^2-\epsilon^2)^2+\epsilon^2\Delta_n^2}
+ \frac{4 f_{1,nj=3/2}\epsilon\Delta_n}
         {(\epsilon_{1,nj=3/2}^2-\epsilon^2)^2+\epsilon^2\Delta_n^2}.
\eeq

For $\epsilon_{1,nj=3/2}<\epsilon<\epsilon_{1,n+1\,j=1/2}$ ($n=2,3,\dots$), we adopted a fitting formula similar to that one used in \cite{rohrmann2018}
\beq\label{aRreson}
\alpha_\text{R}(\epsilon)=\frac{\delta_n}{\beta_n}
 \tan\left[\beta_n\left(\epsilon-\phi_n\right)\right] f(\epsilon),
\eeq
where
\beq \label{xi}
\beta_n = \left\{ \begin{array}{ll}\displaystyle
              \frac{\pi}{2(\phi_n-\epsilon_{1,nj=3/2})} 
                        & (\epsilon_{1,nj=3/2}<\epsilon<\phi_n),\\
     [2ex]\displaystyle 
              \frac{\pi}{2(\epsilon_{1,n+1\,j=1/2}-\phi_n)}
                        & (\phi_n <\epsilon <\epsilon_{1,n+1\,j=1/2}),\\
\end{array} \right.
\eeq
\beq \label{Afit1}
\delta_n = \left\{ \begin{array}{ll}\displaystyle
               315.49655 & (n=2),\\
     [1ex]\displaystyle 
   15.5183449 \times n^{2.9769922}\times (1-A_n)^{-1} & (n>2), \\
\end{array} \right.
\eeq
\beq\label{Efit1}
\phi_n  = \epsilon_{1,n+1\,j=1/2}
 -\left(\epsilon_{1,n+1\,j=1/2}- \epsilon_{1,nj=3/2} \right) B_n,
\eeq
\beq
f(\epsilon)= \left(1 + C_n \xi  \right)
  \left\{ 1 - D_n \left[1-\left(2\xi-1\right)^2 \right] \right\},
\eeq
\beq \label{xi}
 \xi = \left\{ \begin{array}{ll}\displaystyle
              \frac{\phi_n-\epsilon}{\phi_n-\epsilon_{1,nj=3/2}} 
                        & (\epsilon_{1,nj=3/2}<\epsilon<\phi_n),\\
      [2ex]\displaystyle 
             \frac{\epsilon-\phi_n}{\epsilon_{1,n+1\,j=1/2}-\phi_n} 
                        & (\phi_n <\epsilon < \epsilon_{1,n+1\,j=1/2}).\\
\end{array} \right.
\eeq
Quantities $A_n$, $B_n$, $C_n$, and $D_n$ are given as follows:
\beq \label{Afit1}
A_n = \left\{ \begin{array}{ll}\displaystyle
            0.1412(2-\log n)^{2.83}&  (n\le 100),\\
     [1ex]\displaystyle 
            0 & (n>100), \\
\end{array} \right.
\eeq
\beq \label{phi_B}
B_n = \left\{ \begin{array}{ll}\displaystyle
               0.214657809 & (n=2),\\
     [1ex]\displaystyle 
 0.268-10^{-1.2-0.45 (\log n)^2 -1.61\times 10^{-7}(\log n)^{22}} & (n>2).\\
\end{array} \right.
\eeq
For $\epsilon_{1,nj=3/2}<\epsilon<\phi_n$ and $\phi_n <\epsilon <\epsilon_{1,n+1\,j=1/2}$,
\beq \label{cn}
C_n = \left\{ \begin{array}{ll}\displaystyle
        0.7346 - 10^{-0.12 -1.95 \log n +0.035 (\log n)^{-2}}, \\
     [1ex]\displaystyle 
       0.928-10^{-0.35-1.65\log n +0.243(\log n)^{-0.43}}, \\
\end{array} \right.
\eeq
respectively.\ For $\epsilon_{1,nj=3/2}<\epsilon<\phi_n$,
\beq \label{dn2}
D_{n} = \left\{ \begin{array}{ll}\displaystyle
          0.245   & (n=2), \\
     [1ex]\displaystyle 
          0.255   & (3\le n\le 6), \\
     [1ex]\displaystyle 
          0.256 -10^{-1.15-0.22(\log n-0.6)^{-1.12}}  & (n > 6),\\
\end{array} \right.
\eeq
and for $\phi_n <\epsilon < \epsilon_{1,n+1\,j=1/2}$,
\beq \label{dn}
D_n = 0.053-10^{-0.44 -1.965(\log n)^{0.6}}. 
\eeq
The quantity $\phi_n$ represents the energy between two resonances where the real polarizability vanishes. 
Eq. (\ref{phi_B}) gives $B_n$ (the position of $\phi_n$ relative to $\epsilon_{1,n\,j=3/2}$ and $\epsilon_{1,n+1\,j=1/2}$, see Eq. (\ref{Efit1})) with an error $<0.5$\% $\forall$ $n$ and below 0.02\% for $n\ga 30$ (precision increasing with $n$). 
Just outside of resonance cores, where imaginary polarizability is not significant, 
Eq. (\ref{aRreson}) describes the total polarizability redward of $\epsilon_{1,n\,j=1/2}$ with a precision better than $0.9$\% $\forall n$, $0.5$\% for $n>9$ and $0.2$\% for $n>16$. Blueward of $\epsilon_{1,n\,j=3/2}$, the relative error is $1.8$\% for $n=2$, $1.6$\% for $n=3$, $<0.85$\% for $n\ge 4$, and $<0.2$\% for $n\ge 20$.

{\it Postresonance region} ($\epsilon>|\epsilon_{1s_{1/2}}|=1.0000133128$~Ry).
For large values of $\epsilon$, the nonrelativistic Rayleigh cross section converges to the Thomson scattering cross section. This means 
\beq
\left|\alpha_\text{pol}(\epsilon)\right| \rightarrow 4\epsilon^{-2},
\hskip.3in (\epsilon\gg1).
\eeq
A precision better than $0.4\%$ (relative error) for the polarizability above the ionization threshold was obtained with
\beq \label{xi}
\left|\alpha_\text{pol}(\epsilon)\right| = \frac{4}{\epsilon^2}\left[1+
\frac{0.6262}
{1+2.8179 \epsilon^{0.6776} 
 \left(1+0.0216672\epsilon^{1.4745}\right)\log \epsilon}\right].
\eeq
%

\section{Scattering cross section} \label{s:cross}

The Rayleigh scattering cross section was obtained simply by multiplying $|\alpha_\text{pol}|^2$ by the factors appearing in Eq. (\ref{Ra1}). 
\begin{figure}
\includegraphics[width=.48\textwidth]{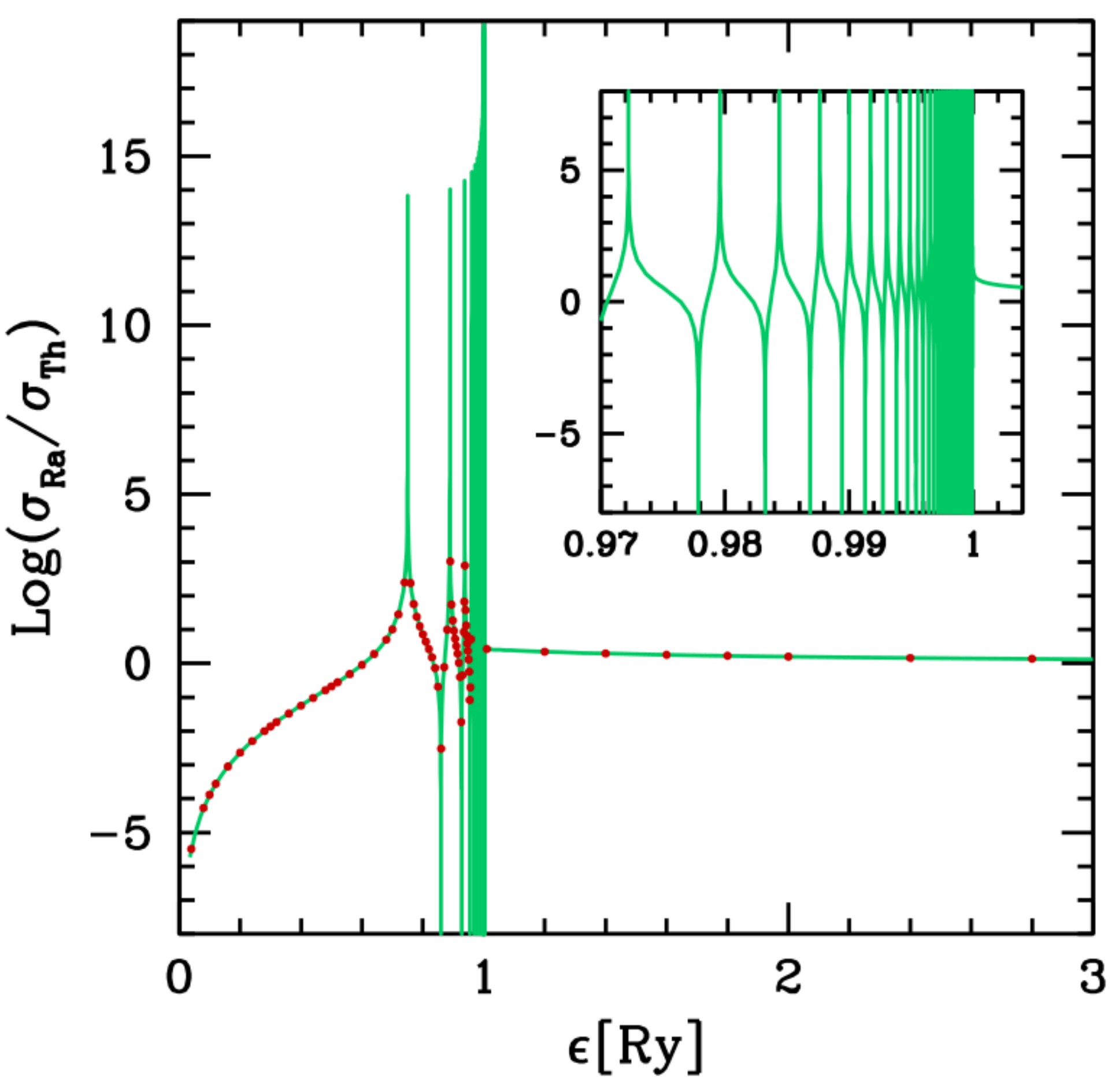}
\caption{Rayleigh cross section in units of the Thompson cross section as a function of the photon energy. The solid line represents evaluations with Eq. (\ref{Ra1}) combined with (\ref{alphaRD}) and (\ref{alphaID}). Symbols represent the results from \citet{gavrila1967}.}
\label{f:fray}
\end{figure}
The Rayleigh cross section for hydrogen atoms obtained in this work is displayed in Fig. \ref{f:fray}. Current calculations (solid line) include about one hundred resonances which have finite amplitudes. 
These results are in very good agreement with those derived from \citet{gavrila1967} in the limited number of energies presented there (symbols), which do not include resonance cores. For high enough energies, in the regime where the dipole approximation still holds, the cross section slowly approaches the expected Thomson formula.

\begin{figure}
\includegraphics[width=.5\textwidth]{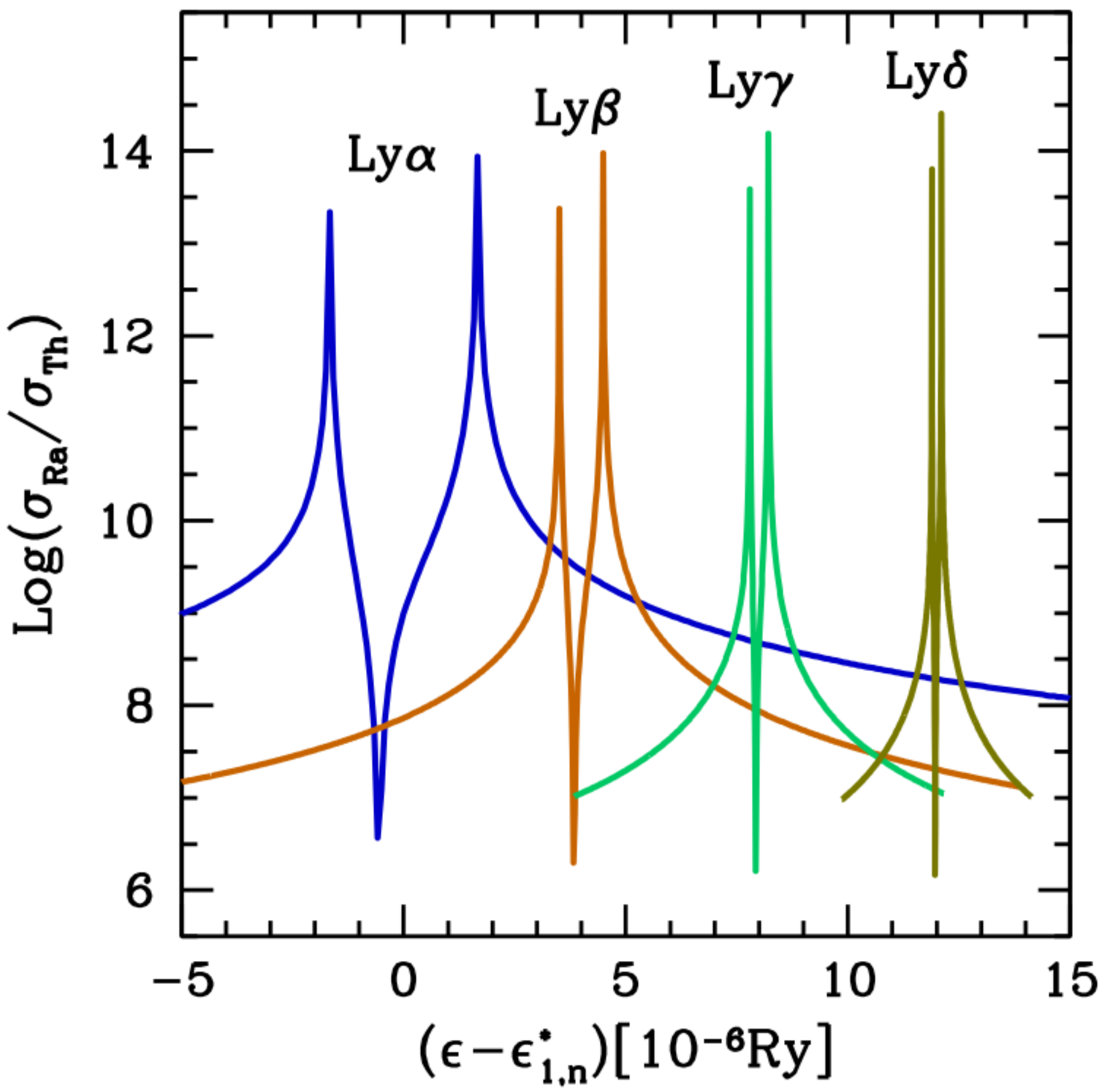}
\caption{Rayleigh cross section for the first four double resonances, from Lyman-$\alpha$ to Lyman-$\delta$, computed with a fine structure and damping effects. Profiles are horizontally offset from each other in steps of $4\times10^{-6}$~Ry.}
\label{f:fLyn}
\end{figure}
It should be noted that the energy interval between successive $n$ states, the separation between fine-structure components, and the natural breadth of the levels scale with $\alpha$ and $n$ in the form
\beq
E_{n+1}-E_n\approx \frac{2}{n^3},\hskip.1in
\Delta E_\text{fine structure}=\frac{\alpha^2}{2n^3},\hskip.1in
\Delta_n \approx \frac{3\alpha^3 \ln(n)}{n^5}.
\eeq
The first of these relations describes the distribution of resonances $1s$~--~$np$ and their accumulation on the photoionization edge (Fig. \ref{f:fray}). The other two relations characterize the shape of each of these (double) resonances, as shown in Fig. \ref{f:fLyn}.
The natural width of the resonances becomes small enough and decreases very quickly as the main quantum number $n$ of the excited state increases faster than the energy separation between fine structure components. Consequently, the profiles of successive resonances are progressively narrower and the magnitude of $|\alpha_\text{pol}|$ in their peaks increases with $n$.  Relative intensities of $1s_{1/2}$--$np_{1/2}$ and $1s_{1/2}$--$np_{3/2}$ resonances are proportional to the ratio 1:2 of their oscillator strengths, which are in turn proportional to the statistical weights of sublevels $np_{1/2}$ and $np_{3/2}$, see Eq. (\ref{f1nj}).

\begin{figure}
\includegraphics[width=.5\textwidth]{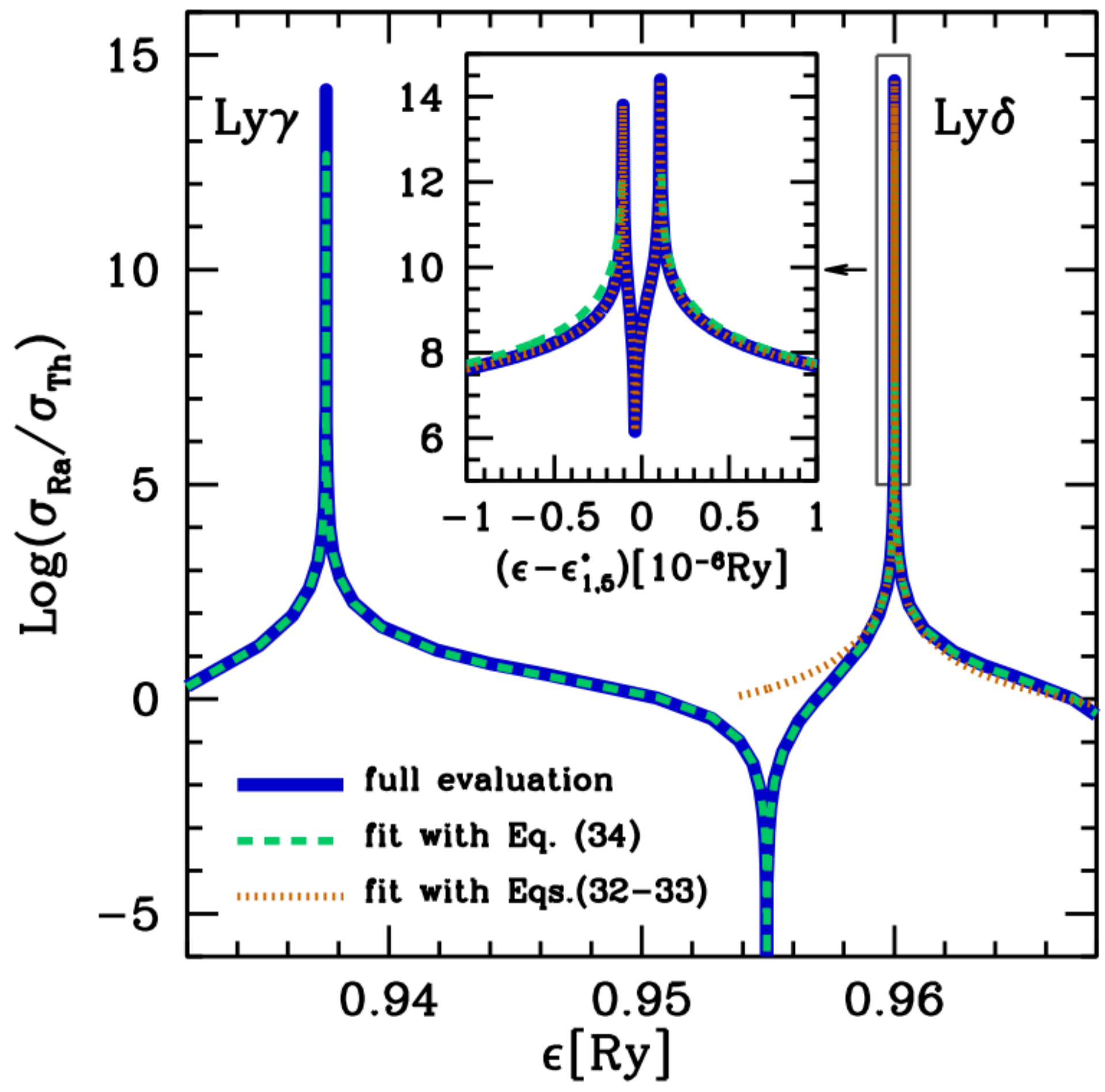}
\caption{Rayleigh cross section for photon energies including the resonances Lyman-$\gamma$ and Lyman-$\delta$. Solid lines represent full solutions based on
Eqs. (\ref{alphaRD}) and (\ref{alphaID}).
Dashed and dotted lines correspond to fitting evaluations with Eqs. (\ref{aRreson}) and (\ref{alphaRDn}-\ref{alphaIDn}), respectively. The inner graph shows the core of  Lyman-$\delta$.}
\label{f:fits}
\end{figure}

As an illustration, Fig. \ref{f:fits} compares the use of polarizability fits in the evaluation of the Rayleigh cross section within the resonance region. Fits based on Eq. (\ref{aRreson}) give satisfactory results where the cross section changes many orders of magnitude over energy intervals between successive resonances. On the other hand, Eqs. and (\ref{alphaRDn}) and (\ref{alphaIDn}) match -- with high accuracy -- the resonance cores including fine-structure details.
In astrophysical conditions where the fine-structure splitting can be considered negligible, Eqs. (\ref{alphaRDn}) and (\ref{alphaIDn}) can be substituted by
\beq\label{a_reson}
\left|\alpha_\text{pol}(\epsilon)\right|= 4f_{1n}
\left[({\epsilon^*_{1,n}}^2-\epsilon^2)^2 + (\epsilon\Delta_n)^2\right]^{-1/2},
\eeq
where $\epsilon^*_{1,n}$ is the mean energy of the transition $1s$~--~$np$.

Current calculations were performed for an isolated atom. It is worth noting that in a realistic plasma, where broadening mechanisms are present due to particle perturbations (collisional broadening) and thermal motions (Doppler broadening), resonance profiles are expected to be significantly broader than those of an isolated radiating atom \citep{omont1972, omont1973, nienhuis1977, burnett1985}. Moreover, interactions with surrounding ions and electrons particularly affect highly excited $np$ states and introduce modifications in the cross section close to the photoionization threshold \citep{griem2005}.

\section{Conclusions}\label{s:concl}

We have performed an accurate numerical evaluation of the Rayleigh scattering cross section for hydrogen atoms in the ground state, including resonances and incident photon energies above the ionization threshold. Current evaluations were carried out using the nonrelativistic dipole approximation in the second-order standard quantum perturbative approach. Due to symmetries of the hydrogen ground state, the calculation can be focused on the atomic polarizability which is expressed in terms of the oscillator strengths' distribution. 
The method is valid for incident photon energies above and below the ionization threshold. It involves a summation over all intermediate electron states which is split into a sum over bound states and a Cauchy principal value integral over the continuum with an imaginary pole term. Convergence in evaluations is achieved by increasing the number of intermediate bound states and quadrature points. 

Our results for Rayleigh scattering are in good agreement with available theoretical data and they expand upon them with a detailed representation of the resonances' region and the incorporation of a fine structure of the bound levels and damping effects due to finite lifetimes of the excited bound states. We provide fitting formulas to obtain the Rayleigh scattering cross sections in the full nonrelativistic domain, as is required for opacity calculations and their use in astrophysical computer codes.

\begin{acknowledgements}
We wish to thank Shigenobu Hirose, who put our attention on current issue. 
We also thank the anonymous referee for the constructive remarks. 
This work was supported by MINCYT (Argentina) through Grant No. PICT 2016-1128. 
\end{acknowledgements}

\bibliographystyle{aa}
\bibliography{HRay}
\end{document}